\documentclass{aa}
\usepackage{latexsym}
\usepackage{graphicx}
\usepackage{natbib}
\usepackage{color}

\definecolor{dgreen}{rgb}{0,0.5,0}

\begin{document}

\title{Disentangling flows in the solar transition region}

\author{P. Zacharias \inst{1,2} \and V.H. Hansteen \inst{1,2} \and J. Leenaarts \inst{3} \and M. Carlsson\inst{1,2} \and B. V. Gudiksen\inst{1,2}}
\institute{Institute of Theoretical Astrophysics, University of Oslo, P.O.Box 1029 Blindern, NO-0315 Oslo, Norway \and Rosseland Centre for Solar Physics, University of Oslo, P.O. Box 1029 Blindern, NO-0315 Oslo, Norway \and Institute for Solar Physics, Department of Astronomy, Stockholm University, AlbaNova University Centre,
SE-106 91 Stockholm, Sweden}

\abstract%
{
The measured average velocities in solar and stellar spectral lines formed at transition region temperatures have been difficult to interpret. 
The dominant redshifts observed in the lower transition region naturally leads to the question of how the upper layers of the solar (and stellar) atmosphere can be maintained. Likewise, no ready explanation has been made for the average blueshifts often found in upper transition region lines. However, realistic three-dimensional radiation magnetohydrodynamics (3D rMHD) models of the solar atmosphere are able to reproduce the observed dominant line shifts and may thus hold the key to resolve these issues.
}{
These new 3D rMHD simulations aim to shed light on how mass flows between the chromosphere and corona and on how the coronal mass is maintained. These simulations give new insights into the coupling of various atmospheric layers and the origin of Doppler shifts in the solar transition region and corona. 
}{
The passive tracer particles, so-called corks, allow the tracking of parcels of plasma over time and thus the study of changes in plasma temperature and velocity not only locally, but also in a co-moving frame. By following the trajectories of the corks, we can investigate mass and energy flows and understand the composition of the observed velocities.
}{
Our findings show that most of the transition region mass is cooling. The preponderance of transition region redshifts in the model can be explained by the higher percentage of downflowing mass in the lower and middle transition region. 
The average upflows in the upper transition region can be explained by a combination of both stronger upflows than downflows and a higher percentage of upflowing mass. 
%
}{
Corks are shown to be an essential tool in 3D rMHD models in order to study mass and energy flows.  
We have shown that most transition region plasma is cooling after having been heated slowly to upper transition region temperatures several minutes before. Downward propagating pressure disturbances are identified as one of the main mechanisms responsible for the observed redshifts at transition region temperatures.
}

\maketitle

\section{Introduction}

Observations of transition region spectral lines reveal the presence of redshifts for lines formed in the temperature range from $\sim$10\,000~K to temperatures of about 2.5$\times 10^5$~K. Above that temperature one often finds that the lines are blueshifted. 
A large amount of literature has been devoted to understanding the observed average Doppler shifts of emission lines formed in the solar transition
region and corona \citep{doschek+al:1976, mariska+al:1978, gebbie+al:1981, feldman+al:1982, dere+al:1984, dere+al:1989, klimchuk:1987, athay+dere:1989, rottman+al:1990, brekke:1993, brekke+al:1997, chae+al:1998, peter+judge:1999, peter:1999, dadashi+al:2011}. These observations have revealed that mean line profiles of transition region lines in the chromospheric network are persistently redshifted by up to 15 km/s in lines formed between $T$ = 0.1 MK to 0.25 MK. These redshifts imply the presence of plasma flows or wave motions in the quiet Sun with amplitudes that are substantial fractions of the sound speed.

A large number of models have been proposed to explain the redshifts in the transition region, including downward propagating compressive waves generated by, for example, nanoflares in the corona \citep{hansteen:1993}, downflows of previously heated spicular material \citep{pneuman+kopp:1977, athay+holzer:1982, athay:1984}, rapid episodic heating at low heights of the upper chromospheric plasma to coronal temperatures \citep{hansteen+al:2010}, and plasma draining from reconnecting loops systems \citep{zacharias+al:2011a}, but no definitive concensus has emerged. 
With the development of realistic three-dimensional radiation magnetohydrodynamics (3D rMHD) models \citep{gudiksen+nordlund:2005, hansteen+carlsson+gudiksen:2007, bingert+peter:2011, gudiksen+al:2011} over the past two decades, detailed comparison between model and observation has become feasible.

\cite{peter+al:2006} found that three-dimensional (3D) numerical models that span the photosphere to corona produce redshifts in transition region lines. In these models, coronal heating is due to the Joule dissipation of currents that are produced as the magnetic field is stressed and braided by photospheric motions. Hansteen et al. (2010) extended this work and showed that transition region redshifts are naturally produced in episodically heated 3D models in which the average volumetric heating scale height lies between the chromospheric pressure scale height of 200 km and the coronal scale height of 50 Mm.

However, even with full access to all relevant variables on a Eulerian grid it has proven difficult to follow the evolution of mass in the outer atmosphere, especially in the transition region where timescales are short, temperature and density gradients are very high and the magnetic topology very complicated. Previously, this problem was tackled by injecting tracer fluids for certain temperature intervals into Bifrost simulations \citep{guerreiro+al:2013} in order to follow mass motions, but naturally, this approach only covers certain regions and not the entire atmosphere and it also proved difficult to follow the dynamics of mass backward in time. The evolution of visible structures in the solar atmosphere, such as fibrils in the chromosphere or magnetic bright points in the photosphere, has been used to follow magnetic field lines and waves over time, in simulations as well as observations \citep{leenaarts+al:2015}. However, \cite{leenaarts+al:2015} concluded that using the swaying motion of fibrils as a tracer of chromospheric transverse oscillations must be done carefully. Magnetic field lines can be traced easily in the simulations by integrating the magnetic field vector from a given grid point for a given distance. However, it is not straight forward to follow magnetic flux as it evolves in time. We have found that the implementation of passive tracer particles, so-called corks, into the Bifrost simulations offers a unique and consistent approach to follow magnetic field lines over time and address the evolution of mass flows between the various layers of the solar atmosphere. 

The structure of the paper is as follows. In Section \ref{simulations}, the rMHD simulation and implementation of the passive tracer particles are described. In Section \ref{methods}, the setup and cork analysis are explained. In Section \ref{section_results}, the results are presented. We finish with a discussion and outlook section (Sections \ref{section_discussion} and \ref{section_outlook}).

\section{Simulations}\label{simulations}

The simulations were performed with the Bifrost stellar atmosphere code \citep{gudiksen+al:2011} in a box extending from the upper convection zone to the low corona. The box size is 24x24~Mm$^2$ in horizontal direction and $\sim$17~Mm in the vertical reaching 2.4~Mm below the solar surface and 14.4~Mm above. The simulation grid has a size of 504x504x496 grid points. This corresponds to a resolution of 48x48~km$^2$ in the horizontal direction. In the vertical direction, a nonequidistant grid spacing was applied and the resolution ranges between 19~km in the photosphere and transition region to $\sim$100~km in the corona, where the temperature and density gradients are less steep.

The Bifrost code includes an equation of state based on solar abundances in local thermal equilibrium (LTE), optically thick radiative transfer, and losses from the photosphere and chromosphere including scattering. In the middle and upper chromosphere and in the transition region, a recipe for effectively thin radiation is used \citep{carlsson+leenaarts:2012}. 
In the corona, thermal conduction along the magnetic field plays an important role. To avoid extremely short time steps, this is treated by operator splitting, where the resulting implicit operator for conduction is solved using a multigrid method. For more details, see \cite{gudiksen+al:2011}.

\begin{figure*}[!h]
\sidecaption
\includegraphics[width=10cm]{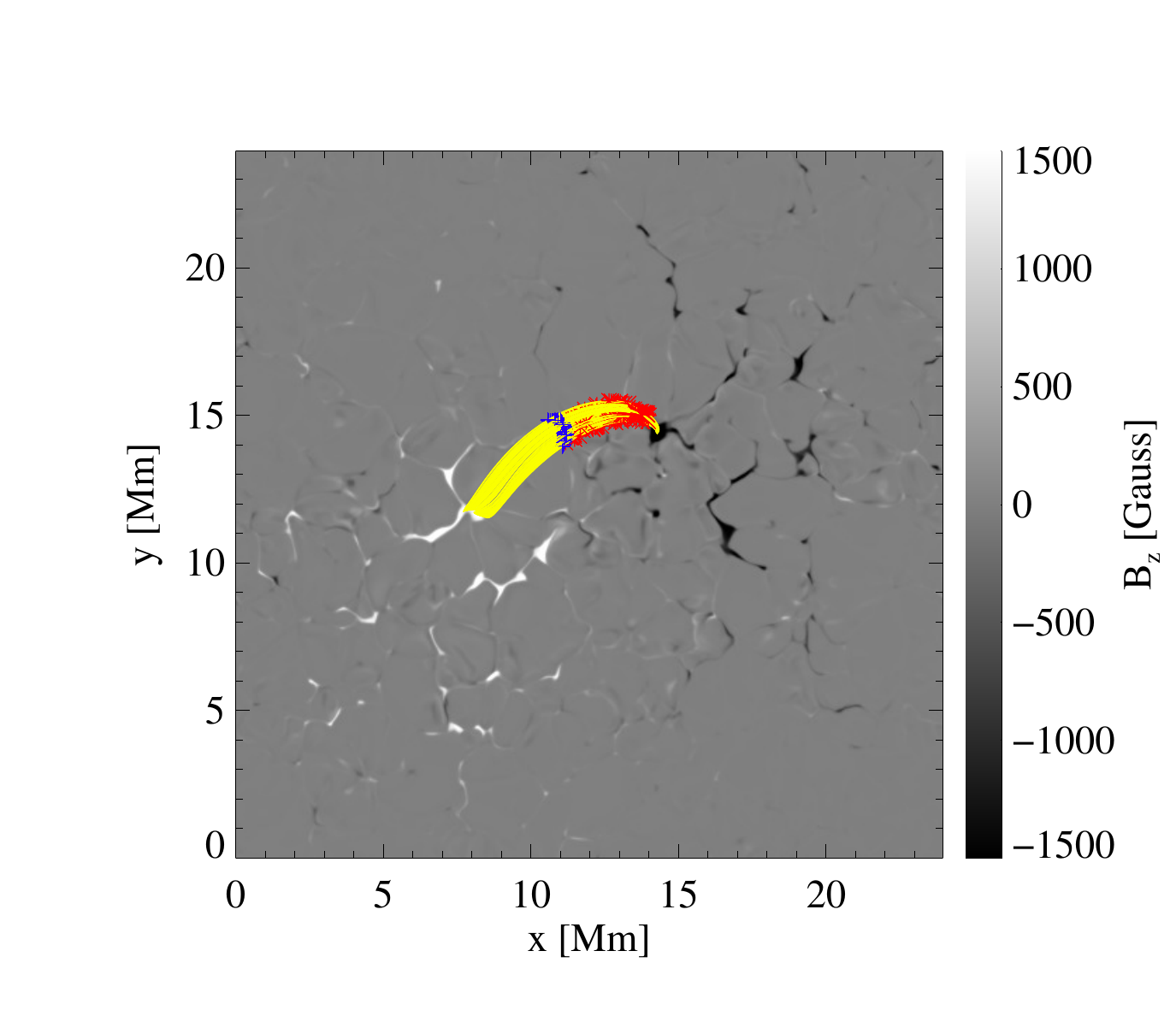}
\includegraphics[width=10cm]{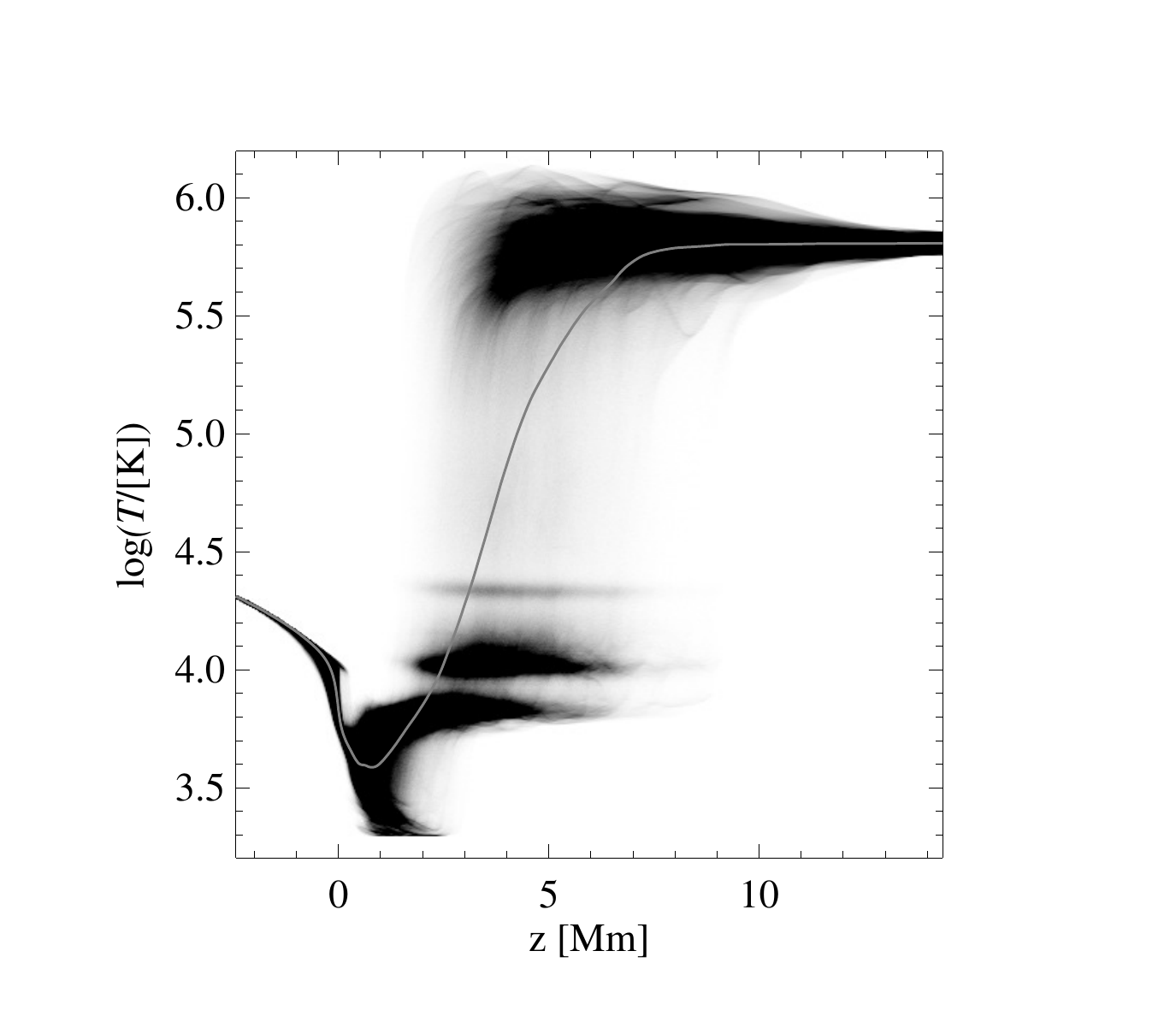}
\caption{{\it{Left panel:}} Vertical magnetic field strength at height $z$=0~Mm for timestep $t$=5000~s. The color bar range is [-1.5 kG, +1.5 kG].  A set of magnetic field lines connecting the main magnetic polarities is overplotted in yellow in the left panel. Field line apices and cork positions of corks with $T=10^5$~K are indicated by blue and red crosses. These field lines are further discussed in section \ref{field_lines_section} {\it{Right panel:}} Temperature distribution of the simulation grid points as a function of height for timestep $t$=5000~s of the simulation. The average temperature at a given height is indicated by the gray line. \label{fig_mag} \label{fig1}}
\end{figure*}

The magnetic field configuration was chosen in a way that it leads to a small network-like configuration. It was created by specifying the magnetic field at the bottom boundary using a potential field extrapolation to compute the field in the entire computational domain. The magnetic field was inserted into a relaxed quasi-steady state hydrodynamical simulation, which is allowed to evolve freely. The simulation was run for 3000~s of solar time using LTE ionization, before nonequilibrium hydrogen ionization was switched on for 830~s. Afterward, starting from $t$=3850~s, the simulation was run for $\sim$1500~s of solar time without nonequilibrium hydrogen ionization including passive tracer particles. More details are provided in \cite{carlsson+al:2016}.

Bifrost is an explicit code with diffusive terms in the equations in order to ensure stability and to ensure that structures smaller than the grid size cannot develop. The diffusive operator employed is split in a small global diffusive term and a location specific hyper diffusion term \citep[see Eq. (9) in][]{gudiksen+al:2011}. Spatial derivatives and the interpolation of variables are performed using high order polynomials. The equations are stepped forward in time using the explicit third order predictor-corrector procedure described by \cite{hyman:1979}.

The simulations are set up in a way to study processes in the solar chromosphere and transition region with a magnetic field configuration characterized as 'enhanced network'. The magnetic field at the bottom boundary consists of two patches of opposite
polarity separated by 8~Mm (see Fig. \ref{fig_mag}, left panel) with an overall balanced flux. The average unsigned magnetic field strength in the photosphere is 48~G (5~mT). The magnetic field distribution does not change significantly during the simulation timespan. Both the top and the bottom boundaries are transparent. At the bottom boundary, the magnetic field is passively advected with no extra field fed into the computational domain. The effective temperature of the simulation is not set directly, but only indirectly by specifying the incoming entropy flux at the lower boundary. The average temperature in this model (Fig. \ref{fig1}, right panel) is maintained by fluid motions in the convection zone, radiative transfer in the photosphere, the balance between acoustic shocks and radiative losses in the lower chromosphere, and the balance of Joule and viscous heating, thermal conduction, and radiative losses in the upper chromosphere, transition region, and corona.\\

The energy equation in the Bifrost code is expressed in terms of the internal energy density per unit volume $e$ as   
\begin{equation}\indent \frac{\partial e}{\partial t} = -\nabla \cdot (e \vec{u}) - p \nabla \cdot \vec{u} + Q_{\rm{rad}} + Q_{\rm{other}} , \label{eq1} \end{equation}
where $\vec{u}$ is the velocity field, $p$ is the gas pressure, $Q_{\rm{rad}}$ is the heating due to radiation, and $Q_{\rm{other}}$ is the heating due to heat conduction, viscous and Ohmic dissipation. 
The various terms contributing to the heating by radiation are 
\begin{equation} \indent Q_{\rm{rad}}=Q_{\rm{phot}} - L_{\rm{chrom}} + Q_{\rm{Ly\alpha}} + Q_{\rm{EUV}}, \label{qrad}\end{equation} 
where $Q_{\rm{phot}}$ is the radiative heating from the photosphere described in \cite{gudiksen+al:2011}, $L_{\rm{chrom}}$ describes losses from the chromosphere due to strong lines, $Q_{\rm{Ly\alpha}}$ is heating from the Ly$\alpha$ line of hydrogen, and $Q_{\rm{EUV}}$ is the heating from extreme ultraviolet (EUV) photons, corresponding to the thin radiative losses from the transition region and corona (negative $Q_{\rm{EUV}}$) absorbed in the chromosphere (positive $Q_{\rm{EUV}}$). The three latter contributions are described in \cite{carlsson+leenaarts:2012}. %
The Joule heating and viscous heating terms are defined as 
\begin{equation} \indent Q_{\rm{joule}} = {\vec{j}}^2/\sigma \label{qjoule}\end{equation} and 
\begin{equation} \indent Q_{\rm{visc}}=2 \eta \xi_{i,j} \xi_{i,j}. \label{qvisc}\end{equation} 
$\vec{j}$ is the current density, $\sigma$ is the electric conductivity, $\eta$ is the viscosity, and $\xi_{i,j}=\frac{1}{2}\left(\frac{\partial u_i}{\partial x_j}+\frac{\partial u_j}{\partial x_i} - \frac{2}{3}\delta_{i,j}\frac{\partial u_m}{\partial x_m}\right)$ is the symmetric part of the viscous stress tensor. 
The thermal conduction term implemented in Bifrost takes the form 
\begin{equation} \indent Q_{\rm{spitz}} = - \nabla F_c,\label{qspitz}\end{equation}
with $F_c = -T^{5/2}\nabla_{\parallel} T$ \citep{spitzer:1956}, where the gradient of $T$ is taken only along the magnetic field and $\kappa_0$ is the thermal conduction coefficient. The conduction
across the field is significantly smaller under the conditions present in the solar atmosphere and is smaller than the numerical diffusion, so it is ignored. 
 The first two terms on the right-hand side of Eq. \ref{eq1} describe the contributions due to the advection of energy and the compression or expansion of the plasma.

\section{Method and analysis}\label{methods}

As a new analysis tool, we have implemented passive tracer particles, so-called corks, into the Bifrost simulations. The tracking of the corks is based on the advection of the corks by the surrounding plasma flows. For the tracking of each cork in the simulation, the plasma flow at the respective cork position is considered for every single timestep. The position $\vec{r}_i$ of a cork labeled $i$ evolves as 
$\frac{\partial \vec{r}_i}{\partial t} = \vec{v}(\vec{r}_i)$, 
with $\vec{v}(\vec{r}_i)$ the plasma velocity at the location of the cork. Initially, corks are placed at every grid point of the simulation cube, which corresponds to a total of 504x504x496=125\,991\,936 individual corks that are followed over time. The snapshot cadence for writing out the basic MHD parameters during the simulation is 10 seconds; we chose the same cadence for writing out the cork positions. This doubles the amount of output data, however, little extra time is needed for the calculation and writing of the additional corks data. The number of corks in the box is assumed to remain constant throughout the simulation. Thus, corks that leave the simulation box on the top or bottom of the box are not lost, but they remain fixed at the same position, which is outside the domain considered for the analysis presented here. Since periodic boundary conditions are applied in the $x$- and $y$-direction, corks that leave the box on either side re-enters the box on the opposite side of the simulation domain.

The main difference between our cork method and the tracer fluid employed by \cite{guerreiro+al:2013} is that the latter is designed to address issues limited to certain temperature regions, whereas the corks cover the entire solar atmosphere without any constraints. The tracer fluid in the work of \cite{guerreiro+al:2013} is placed in a certain volume of the solar atmosphere defined by a specific temperature interval. Inside this specified temperature interval, the plasma density is set to the plasma density of each cell; outside the specified volume, the density is set to $\rho \times 10^{-5}$.
It is therefore much more difficult to follow material in sufficient detail to be able to ask questions such as, where does material currently at 10$^5$ K come from? In addition, the use of corks allows one to study the time evolution of specific magnetic field lines in much greater detail than when using tracer fluids. 
The applications are manifold, as this method allows the tracing of individual plasma structures as well as magnetic field lines backward and forward in time in the most accurate way.

Corks are injected at $t$=3850~s, which is the snapshot that has been used as a reference snapshots in many earlier studies, \citep[{\emph{e.g.,}}][and references therein]{carlsson+al:2016, leenaarts+al:2015}. In particular, our time series is thus suitable for a comparison with the publicly available simulation of an enhanced network region \citep{carlsson+al:2016}.

In principle, many different initial setups for the test particles are possible.  
In this study, the focus is on mass flows throughout the entire solar atmosphere, therefore corks are placed at every single point of the simulation grid. Each cork has a unique ID, which allows us to investigate individual cork tracks and the collective behavior of the corks. In addition, various methods can be applied to follow the corks during the simulation. In this work, we simply follow the corks over time, as they are being advected by the plasma flows at the respective positions in the simulation cube. 

The large number of corks that were tracked during the simulation allow us to perform statistical studies of the collective cork behavior, e.g., plasma flows throughout the entire solar atmosphere, as well as the evolution of various structures, e.g., coronal loops, including their physical parameters such as temperatures, densities, and velocities. 
Extracting the trajectories of the corks is straight forward in this setup. 

In the following, corks with $z$>0 Mm are referred to as atmospheric corks. Initially, 84\% of the corks are atmospheric corks; 25 min later, only 79\% of the corks are located above the surface. At the start of the simulation $\sim$18\% of the atmospheric corks reach the upper boundary ($z$=14.36 Mm) within 25 minutes, while $\sim$6\% reach below the surface. After 25 min solar time, 15\% of all corks have reached the top boundary, and 4\% of all corks have reached the bottom of the box ($z$=-2.44~Mm).

\begin{figure*}[!h]
\begin{center}
\includegraphics[width=0.4\textwidth]{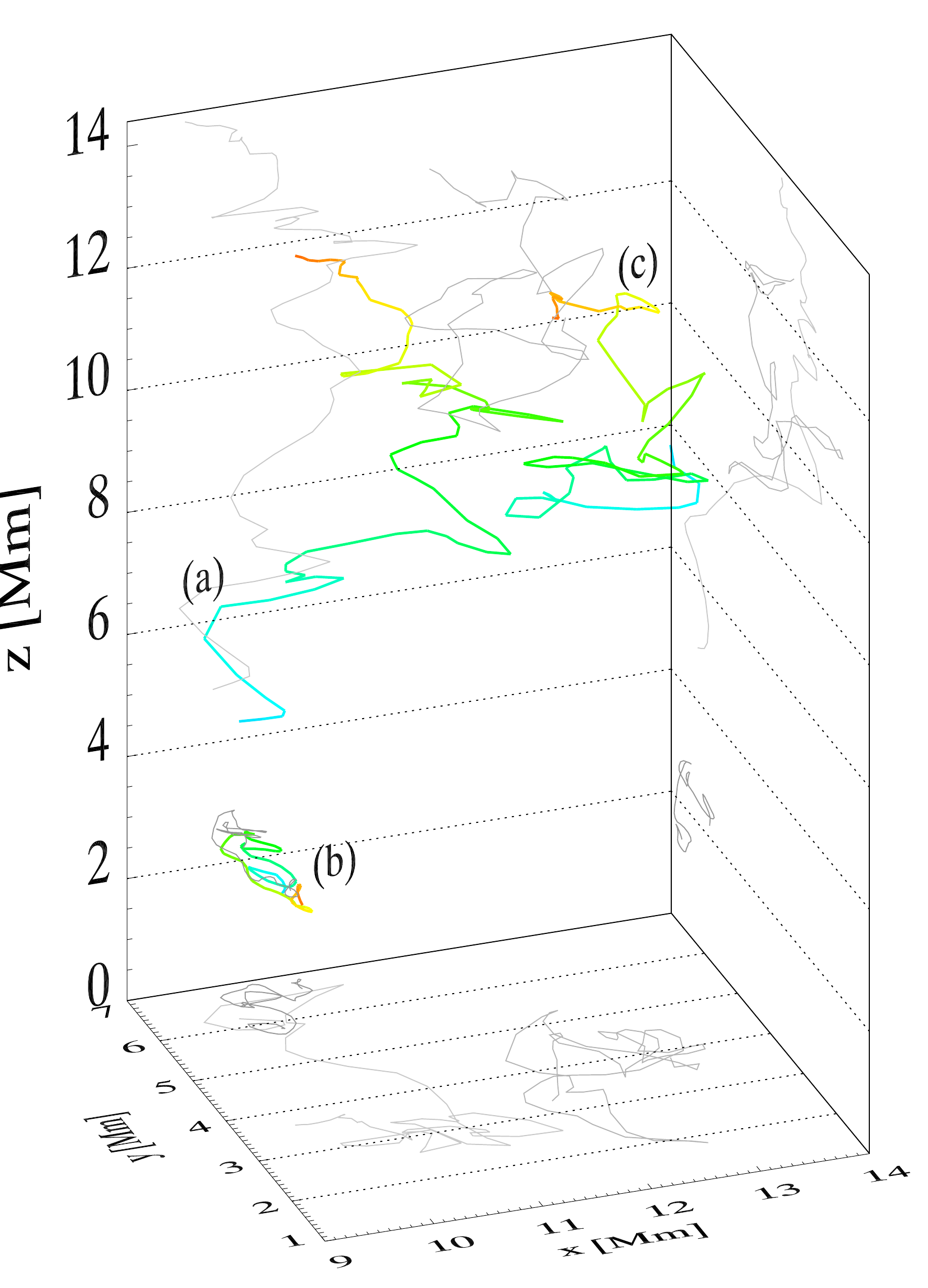}
\hspace{1cm}
\includegraphics[width=0.5\textwidth]{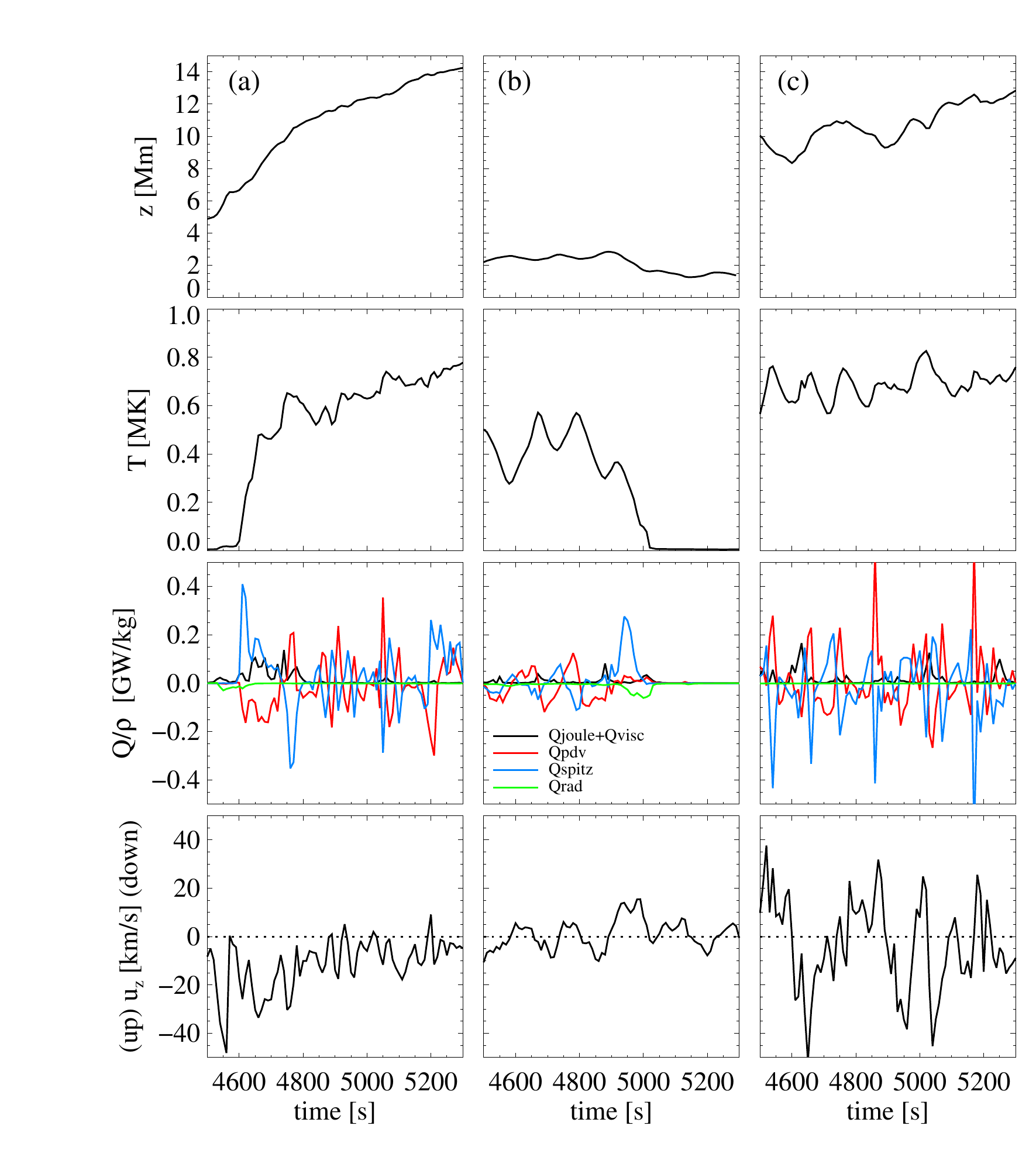}
\end{center}
\caption{{\it{Left panel}}: Cork trajectories of three representative examples described in section 4.1. The temporal evolution for 800~s is shown. As time evolves, the colors change from blue to red. The respective projections in the $xy$-, $xz$- and $yz$-plane are indicated in different shades of gray for the different corks, which have been labeled (a), (b), and (c). {\it{Right panel}}: Shown from top to bottom for the three examples on the left are the temporal evolution of cork height, cork temperature, vertical velocity, and heating and cooling terms per particle, {\emph{i.e.,}} the sum of Joule heating and viscous heating (black), adiabatic compression/expansion (red), Spitzer heat conduction (blue) and radiative losses (green). In the velocity panel, negative velocities are pointing upward, positive velocities are pointing downward. The first column shows the evolution of cork (a), which is heated from chromospheric to upper transition region temperatures starting at $\sim t$=4600~s; the second columns shows the evolution of cork (b), which quickly cools from transition region to chromospheric temperatures at $\sim t$=4900~s; and the third column shows the evolution of cork (c), which remains in the upper transition region throughout the entire simulation. \label{fig2}} 
\end{figure*}

\section{Results}\label{section_results}

In the following, we present results of a numerical experiment, in which we followed corks for approximately 25 minutes of solar time. The corks are initially placed at every grid point of the simulation box. 
At the time that corks are injected into the simulation (at $t$=3850~s), the coronal part of the atmosphere is in a phase of vigorous heating. Thereafter, heating is balanced by energy losses through thermal conduction and radiation, and the average temperature structure remains stable during the second half of the simulation (see Fig. \ref{fig1}, right panel). Compared to the chromosphere and corona, the transition region fills only a very small volume and is characterized by high velocities. Because of these two facts corks do not spend much time in the transition region; hence the population of corks is low in this temperature layer.
In Fig. \ref{fig2}, three examples of cork trajectories are shown, in which corks typical of their starting location have been followed for 800~s following ($t$=4500~s). The evolution of the corks in height, temperature, sum of the Joule heating term and viscous heating term, and vertical velocity is shown on the right.

As an example of a cork that is originally at chromospheric temperatures and then becomes heated to upper transition region temperatures, we consider the cork labeled (a) in Fig. \ref{fig2}.
This cork moves upward in the simulation box, from approximately 5 to 14~Mm, as its temperature rises from $\log T$=3.79 to $\log T$=5.79. We note that the cork traverses the lower transition region temperature range rapidly, in less than $50$~s, following a strong upflow event at $t$=4550~s. The rise in temperature can be associated with a series of small-scale heating events (Joule plus viscous heating, see Eq.s (\ref{qjoule}) and (\ref{qvisc})) taking place over a timespan of $\sim$300~s starting roughly at $t$=4500~s. In addition, the cork experiences strong heating through heat conduction (Eq. \ref{qspitz}) as its temperature increases. This heating balances the cooling due to adiabatic expansion, while the cork is rising in the solar atmosphere. Radiative losses (Eq. \ref{qrad}), on the other hand, are small and can be neglected.
After having reached upper transition region temperatures, there are fewer episodic and burstily occurring Joule heating events and thus the heating rate of the cork decreases. This is to be expected in a model in which the heating is highly episodic and concentrated in certain spatial domains, {\emph{e.g.}}, in current sheets \citep{gudiksen+nordlund:2005, hansteen+al:2015}.

We find that material at lower transition region temperatures typically is either heated (as above), or cools within a short timespan out of the lower transition region temperature range. To illustrate the latter cooling case, we consider a cork that moves at a height of $\sim$2-3~Mm during the first 500~s, while its temperature varies between 300\,000~K-600\,000~K (cork labeled (b)). At $t$=4930~s, the cork temperature quickly drops down to temperatures below 20\,000~K, while the cork moves downward in the solar atmosphere into denser plasma regions. During this period, radiative cooling (Eq. \ref{qrad}) is the most efficient energy term. Preceding this cooling phase, a heating event located on the same loop (see section \ref{field_lines_section}) causes an increase in heat conduction (Eq. \ref{qspitz}) along the loop and a short-term temperature increase between $t$=4900-4930~s. After the cooling phase, the cork remains at low temperature for the rest of the period.

The third and final example shows a cork, which resides in the upper transition region/low corona for the entire period (cork labeled (c)). 
During this time, the cork temperature varies between 600\,000~K and 800\,000~K, as it undergoes a series of episodic heating and cooling events, caused by a combination of Joule and viscous heating (Eq.s (\ref{qjoule}) and (\ref{qvisc})) as well as adiabatic compression and expansion (see Eq. \ref{eq1}) as the cork is moving up and down in the solar atmosphere. During these heating events, the heating rate rises by approximately one order of magnitude compared to the average heating rate. 
The cork remains between 8 and 13 Mm during the entire period and is a typical representative of the coronal and upper transition region corks in this simulation.

In the following subsections, we present the spatial distribution of the corks (section 4.1) and their temperature distribution (section 4.2) in more detail. The analysis of mass flows between different temperature layers and an explanation of the observed transition region redshifts is presented in section 4.3. A scenario explaining the origin of these redshift for a particular set of magnetic field lines is outlined in section 4.4.

\begin{figure*}[!h]
\sidecaption
\includegraphics[width=12cm]{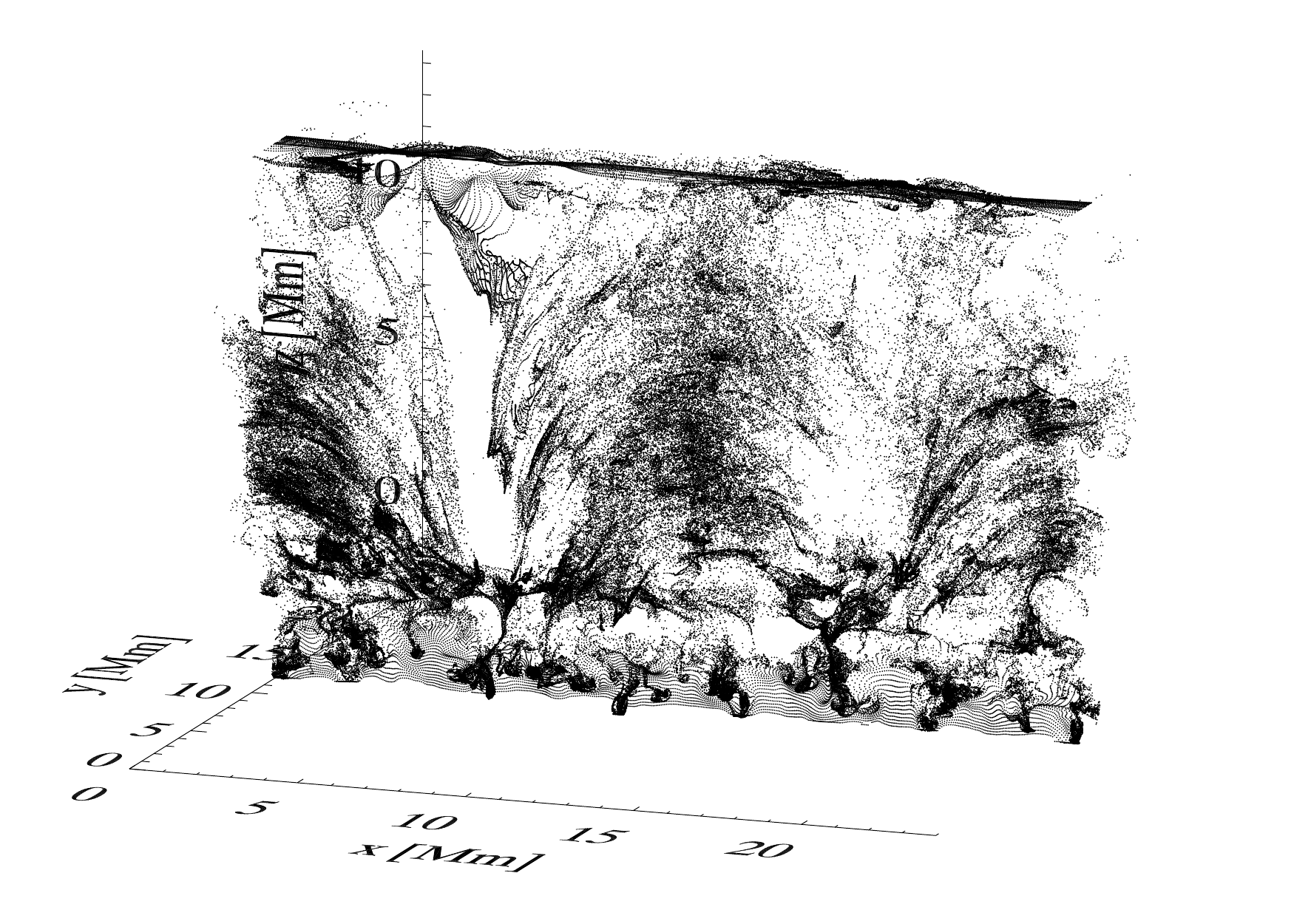}
\caption{Spatial distribution of corks, which have initially been placed in the $y$=12~Mm plane of the simulation box 15 minutes into the simulation. Every single dot indicates the position of one cork. Dark regions indicate regions, where the cork density is high; white regions represent voids of corks, {\emph{i.e.,}} regions where no corks are present. As time evolves, the corks start moving in both horizontal and vertical direction, thus slowly moving out of the $y$=12~Mm plane (see section \ref{overall}). A movie showing the evolution over 25 minutes is available in the online edition. The movie is also available at http://folk.uio.no/piaz/papers/corks.avi.}
\label{fig3}
\end{figure*}

\subsection{Overall distribution of the corks}\label{overall}

Figure \ref{fig3} shows a snapshot of the ($x,y,z$)-positions of the corks 15 minutes after their injection. The corks were initially placed in the $xz$-plane at $y$=12~Mm. A movie of the entire time series can be found in the online material. Every single dot indicates the position of one cork. Dark regions indicate regions of high cork density; white regions show regions where no corks are present. Very quickly, the corks spread throughout the simulation box and start outlining different morphological structures, including the photospheric granulation pattern close to the $z$=0~Mm plane, large loop structures connecting the two main magnetic polarities located at approximately $x$=7~Mm and $x$=15~Mm, and short, low-lying loop-like structures. After a few minutes, much fewer corks are found above the magnetic field concentrations compared to more quiet magnetic regions (see Fig. \ref{fig3} and Fig. \ref{fig8}, left panel). These voids are due to flow divergence and strong plasma flows above the strong magnetic field regions. In the following, we focus mostly on the vertical motion of the corks, but one should note that the corks are moving in the horizontal directions as well, as indicated by the projections of the trajectories in Fig. \ref{fig2}.

As the simulation progresses, corks rapidly distribute in height throughout the box. 
Depending on the timestep, it takes between 3~min and 6~min until the corks with an initial height of 5~Mm have traveled all the way up to the top of the box, about 5-8~min until corks with initial height of 3~Mm reach the top and 15-21~min for corks with an initial height of 1.5~Mm to reach that high. On the other hand, none of the corks with an initial height of 5~Mm ever reach below the surface within the time sequence of 25~min we analyzed. It takes between 20~min or longer for the corks with an initial height of 3~Mm to reach the surface and 7-19~min for the corks with an initial height of 1.5~Mm. Thus, the downward moving corks take much longer to reach the surface than the upward moving corks take to reach the top.

\begin{figure*}[!h]
\includegraphics[width=0.45\textwidth]{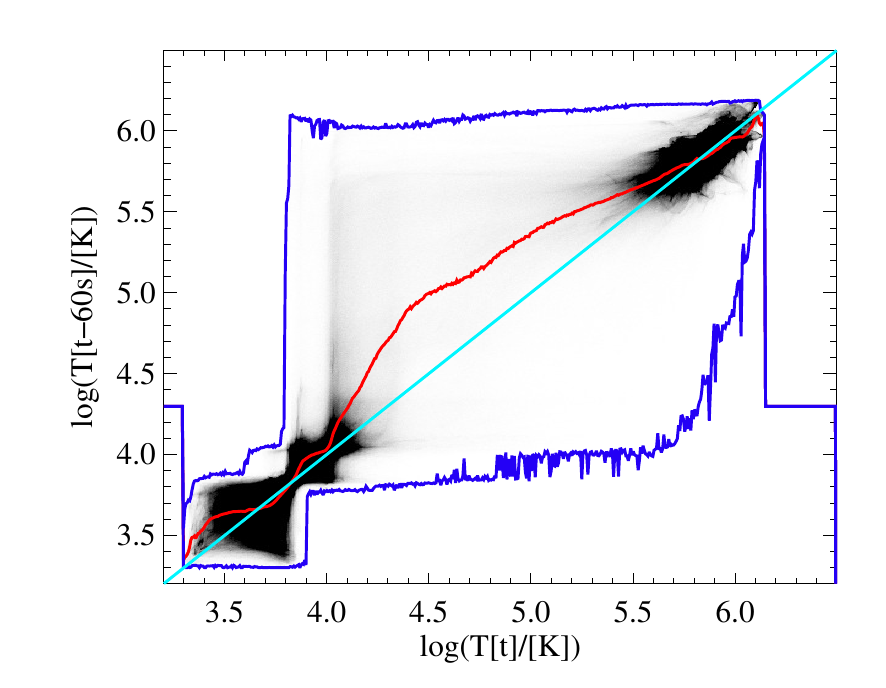}
\includegraphics[width=0.45\textwidth]{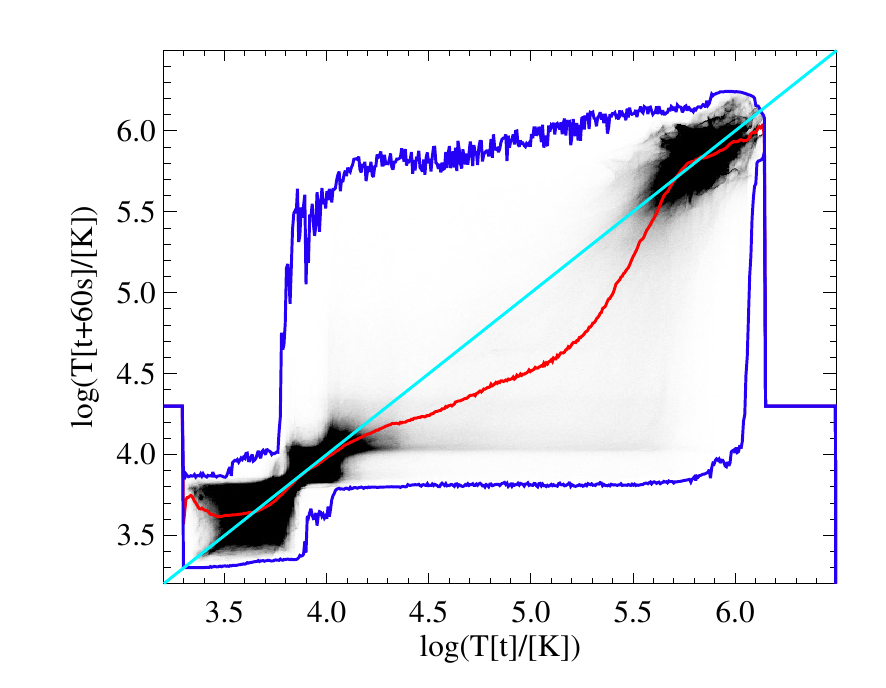}
\caption{Probability density functions (PDFs) of cork temperature. The PDFs are shown with respect to $t$=5000~s for $t$=4940~s (left panel) and $t$=5060~s (right panel) (see section \ref{where}). The distribution of corks at $t$=5000~s is shown by a turquoise line in each panel. The average cork temperature at each timestep is indicated by the red line. The blue lines show the minimum and maximum temperature of all corks at the respective height.}
\label{fig4}
\end{figure*}

\subsection{Temperature evolution of transition region corks}\label{where}
The evolution of mass flows is most complicated for mass at transition region temperatures, hence to clarify the mass cycle for this gas we concentrate explicitly on the details of this temperature range. 
Perhaps the most striking development when looking at the evolution of the temperature distribution of the corks is the large drop in average temperature over large parts of the transition region temperature range within a short time. In Fig. \ref{fig4}, probability density functions (PDFs) of cork temperatures are presented for $t$=4940~s (left panel) and $t$=5060~s (right panel). The average temperature is shown by the red line, and values of the reference timestep $t$=5000~s are indicated by a turquoise line. Around $T$=100\,000~K, the drop in average temperature within two minutes amounts to an order of magnitude.

\begin{figure*}[!h]
\includegraphics[width=0.95\textwidth]{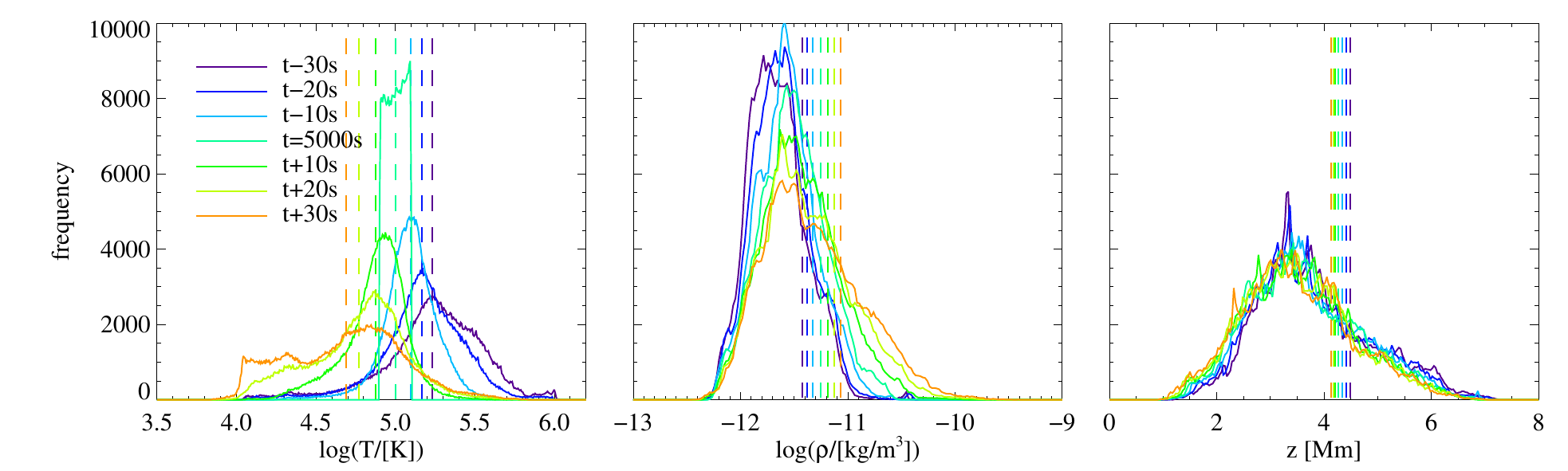}
\caption{One-minute evolution of temperature, density, and height distributions of corks in the temperature range $\log (T/[K])$=[4.9,5.1] at $t$=5000s. The  evolution between $t$-30s and $t$+30s is shown by the different colors. The respective mean values are indicated by dashed lines. \label{fig9}}
\end{figure*}

Following a similar approach as \cite{guerreiro+al:2013}, we investigated the temperature evolution of corks at transition region temperatures in more detail by following all corks in the temperature range $\log(T/[K])=[4.9,5.1]$ over time for several minutes. Because our cork-based approach allows for both backward and forward tracing, we can thus also draw conclusions on the situation prior to an observed state.
Figure~\ref{fig9} shows the temperature, density, and height evolution of corks in the temperature range $\log(T/[K])=[4.9,5.1]$ at $t=5000$~s for $\pm$30~s (we have considered several other timespans with similar results).  
We find that after one minute these corks have cooled down from an average temperature of $\sim 150\,000$~K to $\sim 50\,000$~K. At the same time, the average mass density of the corks increases by roughly half an order of magnitude and the average height decreases by $\sim 0.5$~Mm indicating that, on average, the corks are sinking lower into the atmosphere. The presence of global oscillations in the simulation complicates this analysis for certain timesteps, but on average, the same trend is observed for the entire time series.

After two minutes of solar time, only 4\% of the corks are still in the same temperature range ($\log T=[4.9,5.1]$), whereas 86\% have cooled to lower temperatures and 64\% all the way down to below 20\,000K (see Table~\ref{times}). At the same time, only 10\% of the corks have been heated, {\emph{i.e.}}, around 4\% to temperatures above $300\,000$~K. Only very few corks, around $0.02$\%, reach temperatures above $\log(T/\mbox{[K]})=6.0$. Similar numbers are found for various other timesteps, suggesting a fast cooling process for the transition region corks once they reach the $T=100\,000$~K temperature range. Our results are in agreement with those of \cite{guerreiro+al:2013} who injected a tracer fluid filling the volume between $20\,000$~K and $300\,000$~K into the Bifrost simulations. After two minutes of solar time, they found that 95\% of the material has left this region, about 70\% of which has cooled to chromospheric temperatures, while 25\% of the material was heated to higher upper transition region or coronal temperatures. The remaining 5\% of the material remained within the initial temperature range.

On the other hand, in contrast to what was speculated by \cite{guerreiro+al:2013}, we find no evidence for rapid heating to coronal temperatures shortly before the cooling phase sets in. Two minutes prior to $t=5000$~s, the vast majority of the corks ({\it i.e.,} 85\%) already exist at temperatures above $100\,000$~K, while 4\% are within the same temperature interval, whereas only 11\% have lower temperature. Indeed, most of the corks already have temperatures higher than $100\,000$~K as early as 10 minutes before they transit the $T=100\,000$~K interval.

\begin{table}
\begin{center}
\caption{Temperature analysis for corks with $\log T$=[4.9,5.1] at $t$=5000~s. Percentages of corks within a given temperature interval are stated for various timesteps before and after (see section \ref{where}). \label{times}}
\begin{tabular}{lrrr}
\hline
\hline
        & $\log T$<4.9 & $\log T$=[4.9,5.1]  & $\log T$>5.1 \\ 
             \hline
-10 min & 6  &   2    &  92 \\
-2 min   & 11 &  4    &  85 \\
-1 min   &  11 &  7    &  82 \\
-30s      &  12 & 15   &  73 \\
$t$=5000s & - & 100 & - \\
+30s     &  70 & 18   &  11 \\
+1 min  &  79 & 10   &  11 \\
+2 min  &  86 &  4    & 10 \\
\hline
\end{tabular}
\end{center}
\end{table}

\subsection{Mass flows in the solar atmosphere}\label{section_flows}

\begin{figure*}[!h]
\includegraphics[width=0.95\textwidth]{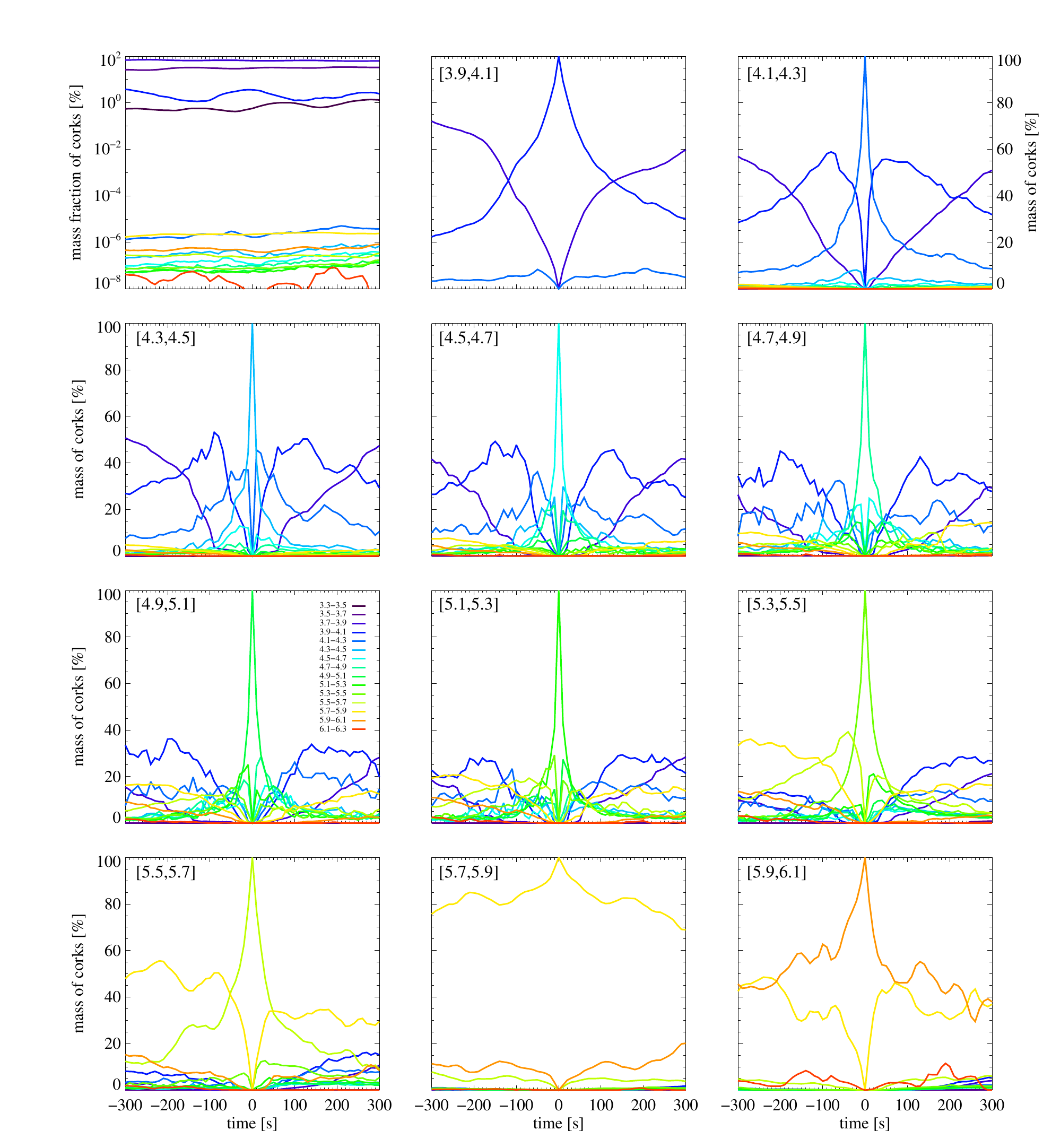}
\caption{ {\it{Upper left panel}}: Temporal evolution of cork masses in a given temperature bin for a 10-minute timespan around $t$=5000~s. Only atmospheric corks are considered. The temperature bins are divided into temperature intervals of $\log T$=0.2~dex, and each bin is represented by a different color (see legend in panel labeled [4.9,5.1]). The fraction with respect to the total mass of all atmospheric corks is shown (see section \ref{section_flows}). {\it{Rows 1 (middle and right panel)-4:}} Temporal evolution ($t\pm$300~s) of the mass of corks within a given temperature bin. Only corks that contribute to the selected bin are considered, and all panels are normalized to 100. The same legend applies for all panels. \label{fig5}}
\end{figure*}

The mass assigned to each cork is determined from the particle mass density times the volume of the cell that the cork is in at $t$=3850~s. Throughout the entire simulation, the mass represented by each cork is assumed to be constant. Because of the continuous flows between the lower and upper layers of the solar atmosphere, the number of corks in each temperature interval varies with time, and the mass within each temperature layer is changing constantly.
In order to disentangle the flows between different atmospheric layers, the corks were divided into various $\log T$ intervals ranging from $\log(T/[K])$=3.3 to 6.3 at intervals of 0.2 dex in $\log T$.
Owing to the density stratification of the solar atmosphere, corks with high masses are overwhelmingly found at low temperature (see Fig. \ref{fig5}, upper left panel). 

In the following, the mass evolution of corks in different temperature bins is investigated. Results are presented in Fig. \ref{fig5} and described in more detail in the following. Only the contributions to each selected bin are considered and all panels have been normalized to 100. As a representative timestep, $t$=5000~s is chosen, and the mass evolution of the corks is investigated for $\pm$ 300~s. The arrangement of the panels is of ascending order in temperature.

Most plasma at low transition region temperatures is shown to be episodically cycling between chromospheric and transition region temperatures. A small fraction of the mass ($\sim$3\%) is found to have its origin at upper transition region temperatures, cooling to low transition region temperatures on a timescale of a few minutes; similarly a small fraction ($\sim$2\%) is reheated to upper transition region temperatures within five minutes. 

 As we move to higher temperature channels, more temperature channels contribute in terms of mass. The contribution from lower temperature channels dominates, however, the fraction of hot mass that is cooling and cool mass that is being heated to upper transition region temperatures increases constantly. 
 In the low and middle transition region (Fig. \ref{fig5}, second row), mass exchange takes place on relatively short timescales and the contributions of the various temperature channels are rather symmetric.
 
In the middle transition region temperature intervals $\log T=[4.9,5.1]$, [5.1,5.3], and $[5.3,5.5]$ (Fig. \ref{fig5}, third row), the asymmetry of the lines indicates that the fraction of hot mass that cools is higher than the fraction of cool mass that gets heated. This imbalance grows with higher temperature. On average, mass is cooling as it transits these bins. The temperature interval $\log T=[5.3,5.5]$ is the temperature range, where a significant amount of material is cooling. On timescales of ten minutes, a large percentage of material is found to cool all the way from $\log T$=[5.7,5.9] to $\log T$=[3.9,4.1], after passing through the $\log T$=[5.3,5.5] temperature band. 

As we move up higher in the solar atmosphere, the distributions of cork masses become symmetric again, and the heating and cooling processes happen on longer timescales of several minutes. Most mass at upper transition region temperatures, {\it i.e.}, $\log T=[5.7,5.9]$ is already found in the same temperature interval several minutes earlier, and remains at the same temperature for several minutes after. The other half of the material has originally been heated from the temperature channel just below and some 10\% of the mass has cooled from coronal to upper transition region temperatures within five minutes. A small amount of hot plasma ($\sim$2\%) cools to temperatures of $\log T=[3.9,4.3]$ within five minutes. 

In the corona, about half of the mass in the temperature range $\log T=[5.9,6.1]$ is heated from $\log T$=[5.7,5.9] on timescales of a few minutes. The other half of the material is found to remain at coronal temperatures for several minutes before and after. Slightly less than half of the mass cools to the $\log T=[5.7,5.9]$ interval afterward and about half of the mass remains at coronal temperatures, however, a fraction of hot material ($\sim$8\%) is found to cool all the way down to chromospheric and lower transition region temperatures ($\log T=[3.9,4.3]$).


In the following, density-squared-weighted vertical velocities are considered for the various temperature layers discussed above. These velocities are of interest, since they are proportional to the intensity of optically thin emission lines and thus to the observed Doppler shifts ($\rho^2 \propto I \propto v_D$). To reduce the effects of global oscillations present in the simulation, we analyze temporal averages of the vertical velocities for the timesteps $t$>4500~s, when the simulation has settled down. 
 The chromosphere is permeated by (mainly) upwardly propagating acoustic/slow mode waves and shocks. Therefore, we rather concentrate on average flows in the transition region and corona, {\it i.e.}, on plasma with temperatures above $\log T$= 3.9. 
 
\begin{figure*}[!h]
\begin{center}
\includegraphics[width=0.95\textwidth]{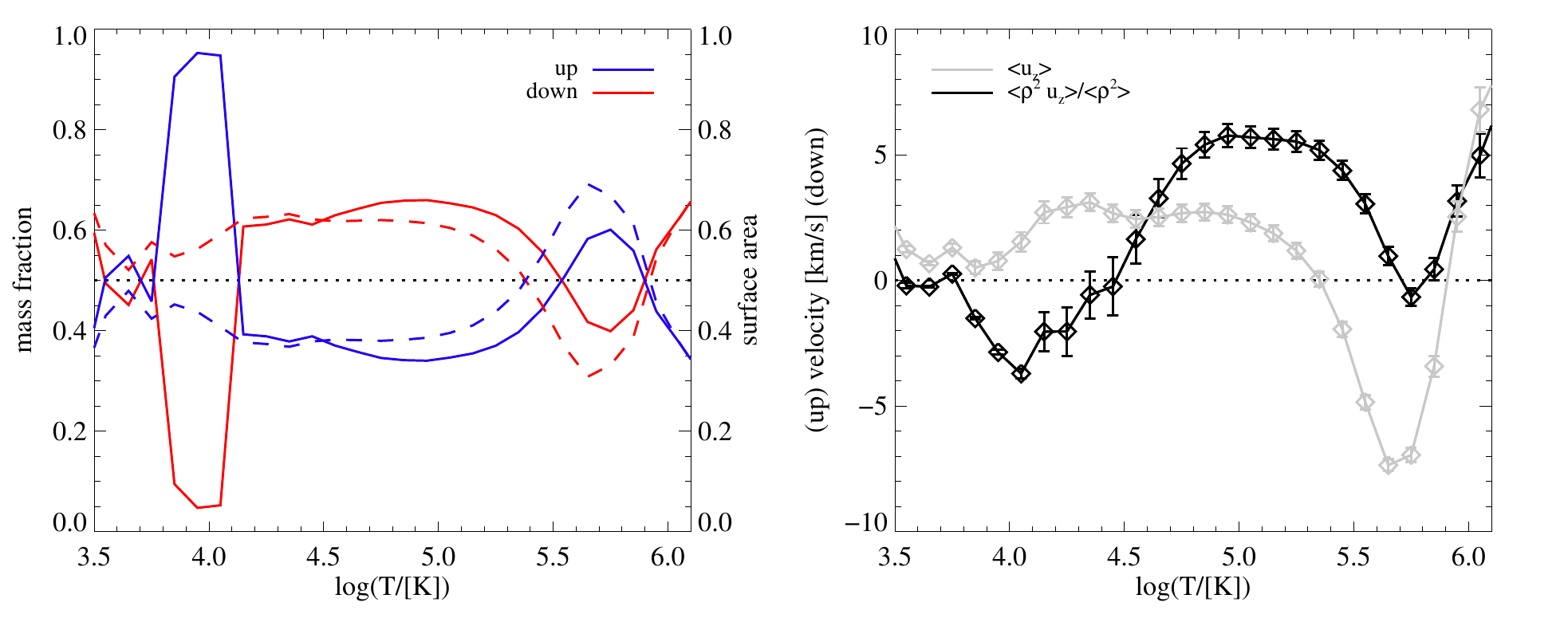}
\end{center}
\caption{{\it{Left panel:}} Mass fraction (solid lines) and surface area (dashed lines) of upflowing (blue) and downflowing (red) plasma in various temperature intervals in the range $\log T$=[3.5,6.1]. Temporal averages based on the timesteps between $t$=4500~s and $t$=5300~s are presented. {\it{Right panel:}} Average vertical velocity (light gray line) and density-squared-weighted vertical velocity (black line) as a function of temperature. The latter represents the trend of the average Doppler shift with line formation temperature. The error bars show the mean error. \label{fig7}}
\end{figure*}

In the right panel of Fig. \ref{fig7}, the density-squared-weighted vertical velocity (black line) is shown together with the average vertical velocity (light gray line) as a function of temperature.  On the left, the respective mass fractions of upward (blue) and downward (red) moving plasma are shown (solid lines) together with the respective fractions of surface area (dashed lines). This analysis is based on the actual MHD variables to avoid uncertainties due to uneven cork distributions. However, a comparison between corks and MHD variables shows that similar results are obtained for both.  
On average, more mass is upflowing than downflowing in the upper chromosphere, {\emph{i.e.,}} in the temperature interval $\log T$=[3.9,4.1]. This large percentage of mass is due to spatially localized upflows of high density plasma. Even though the surface area containing downflows is larger than surface area containing upflows, and the average vertical velocity in this temperature range is slightly pointing downward, these high density upflows result in observed blueshifts in the model.

Throughout the low and middle transition region, {\it i.e.}, $\log T=[4.1,5.5]$, the fraction of downflowing mass outweighs the fraction of upflowing mass by $\sim$20\%, and the surface area with downflows is larger than the surface area with upflows in the temperature range $\log T=[4.1,5.3]$. The average vertical velocity is pointing downward, and since the average density of the plasma that is downflowing is higher than the average density of the plasma that is upflowing, strong average redshifts are observed. A similar situation (higher percentage of downflowing than upflowing mass) is found in the corona ($\log T$=[5.9,6.3]). On the other hand, in the upper transition region ($\log T$=[5.5,5.9]), the situation is reversed. The fraction of upflowing mass exceeds the fraction of downflowing mass, and the surface area covered by upflows is larger than the surface area covered by downflows. However, the average density is lower in regions with upflows than in regions with downflows. Therefore, the strong average upflows observed at upper transition region temperatures do not carry as much weight, and thus the average blueshift ($\propto \rho^2$) observed in this model is notably smaller than the average upflow velocity.

In summary, the average density-squared weighted velocity, which reflects the trend of the average Doppler shifts with line formation temperatures, is pointing downward in the lower and middle transition region, because the fraction of downflowing mass outweighs the fraction of upflowing mass. This explains the observed transition region redshifts in the model. The situation is reversed in the upper transition region, where we find that more material is upflowing than downflowing. 
Furthermore, in addition to having a larger mass fraction, downflows at transition region temperatures are found to last longer than upflows. For example, at $t=5000$~s, 25\% of the corks at $\log T$=[4.9,5.1] show downflows for one minute or longer, while only 2\% show upflows for one minute or longer. 
On average, $\sim$20\% of the corks show downflows for one minute or longer, while only 7\% show upflows for one minute or more. However, we note that material that is upflowing or downflowing for a minute or longer, no longer contributes to the $\log T$=[4.9,5.1] interval, since it is likely to be hotter or cooler after some 30 seconds.

\subsection{Flows along magnetic field lines}\label{field_lines_section}
So far, we only considered vertical (mass) flows. Since plasma motions and thermal conduction are constrained to flow along the magnetic field in a low-$\beta$ plasma, we focus on plasma flows {\it{along}} magnetic field lines in this section.

We again concentrate on typical corks, which are initially at mid-transition region temperatures, as the dynamics of these corks has proven to be the most complicated. 
Magnetic field lines have been traced through all 27\,246 corks in the temperature range $\log(T/[K])=[4.99,5.01]$ at $t=5000$~s. A very small fraction, 0.2\%, of those field lines do not connect back to the photosphere and have only one footpoint rooted in the photosphere. These {\it{open}} field lines are located in strong network regions. 

\begin{figure*}[!h]
\sidecaption
\includegraphics[width=12cm]{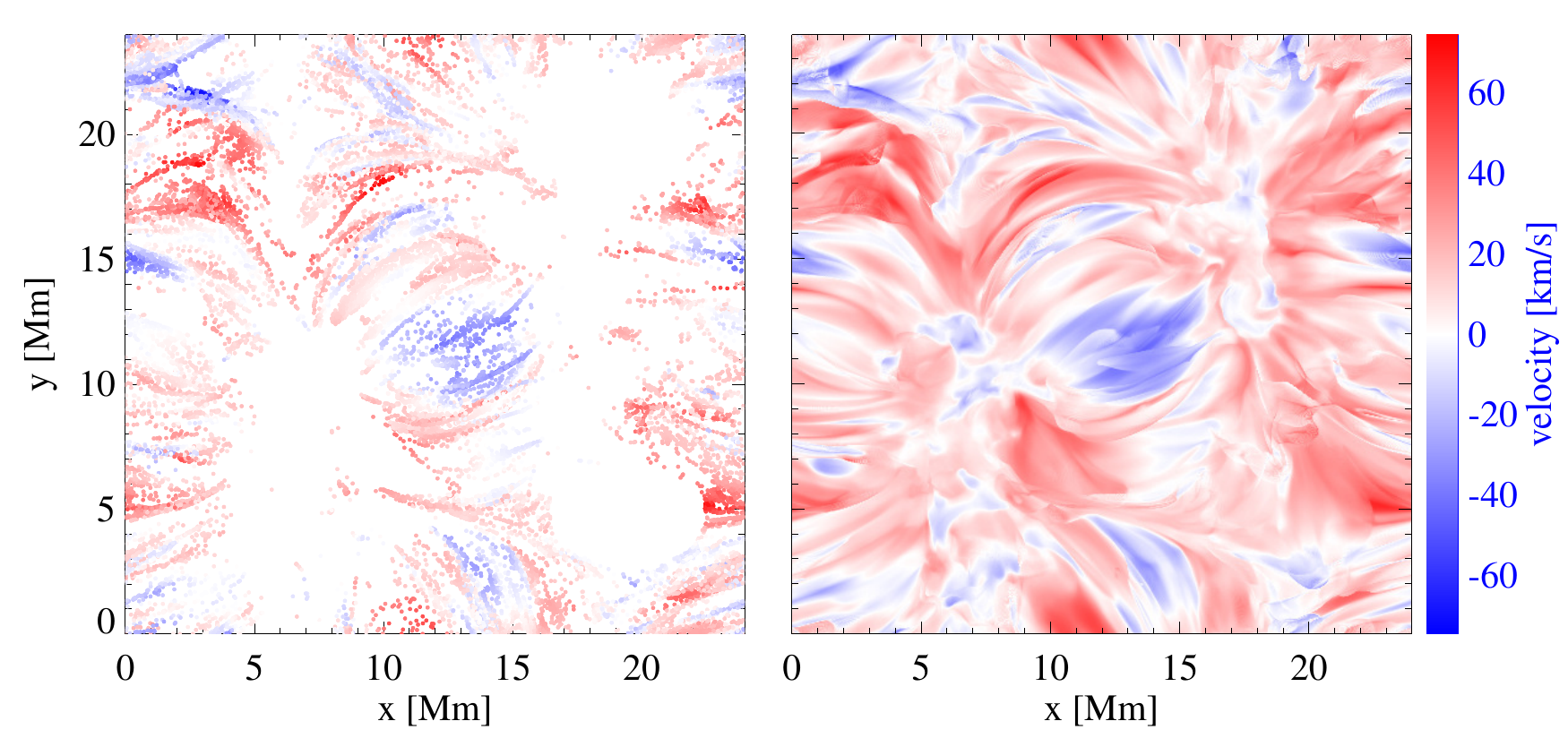}
\caption{{\it{Left}}: $xy$-positions of corks with $\log T$=[4.99,5.01] at $t$=5000s color coded with vertical cork velocity. Red indicates downflows; blue shows upflows. {\it{Right:}} C\,{\sc{iv}} (1548\AA) Doppler shift map for $t$=5000s. Blue indicates blueshifts (upflows), red shows redshifts (downflows). The color bar on the right shows the velocity scales; the same range has been chosen for both panels. \label{fig8}}
\end{figure*}

Roughly 70\% of the corks at $T=100\,000$~K show downflows, $\sim$30\% show upflows, and the average vertical velocity is found to be $\langle u_z\rangle$=6.7~km/s. Their spatial distribution is shown in the left panel of Fig. \ref{fig8}; the color coding represents the vertical velocity of each cork. The cork velocity map bears close resemblance to the C\,{\sc{iv}} Doppler shift map, in regions where corks are present. A synthesized Doppler shift map of the C\,{\sc{iv}} (1548\AA) line formed at $T\sim 100\,000$~K is shown in Fig. \ref{fig8} (right panel). The average Doppler velocity is found to be 8.9~km/s at $t=5000$~s. In the strong network regions, where strong upflows dominate, very few corks with $T$=$100\,000$~K are found.

The cork heights are distributed between 1.3~Mm and 7.8~Mm with a mean of 4.1~Mm. The loop apices are distributed between 2.5~Mm and 13.2~Mm with a mean of 4.8~Mm. Only about 1\% of the corks lie on field lines with apex heights greater than 8~Mm. Both cork heights and apex heights are slightly lower for the downflowing corks as compared to those that are upflowing. 
Most field lines are roughly aligned in $x$-direction, which is due to the underlying magnetic field configuration (see Fig. \ref{fig1}, left panel).

\begin{figure*}[!h]
\begin{center}
\includegraphics[width=0.8\textwidth]{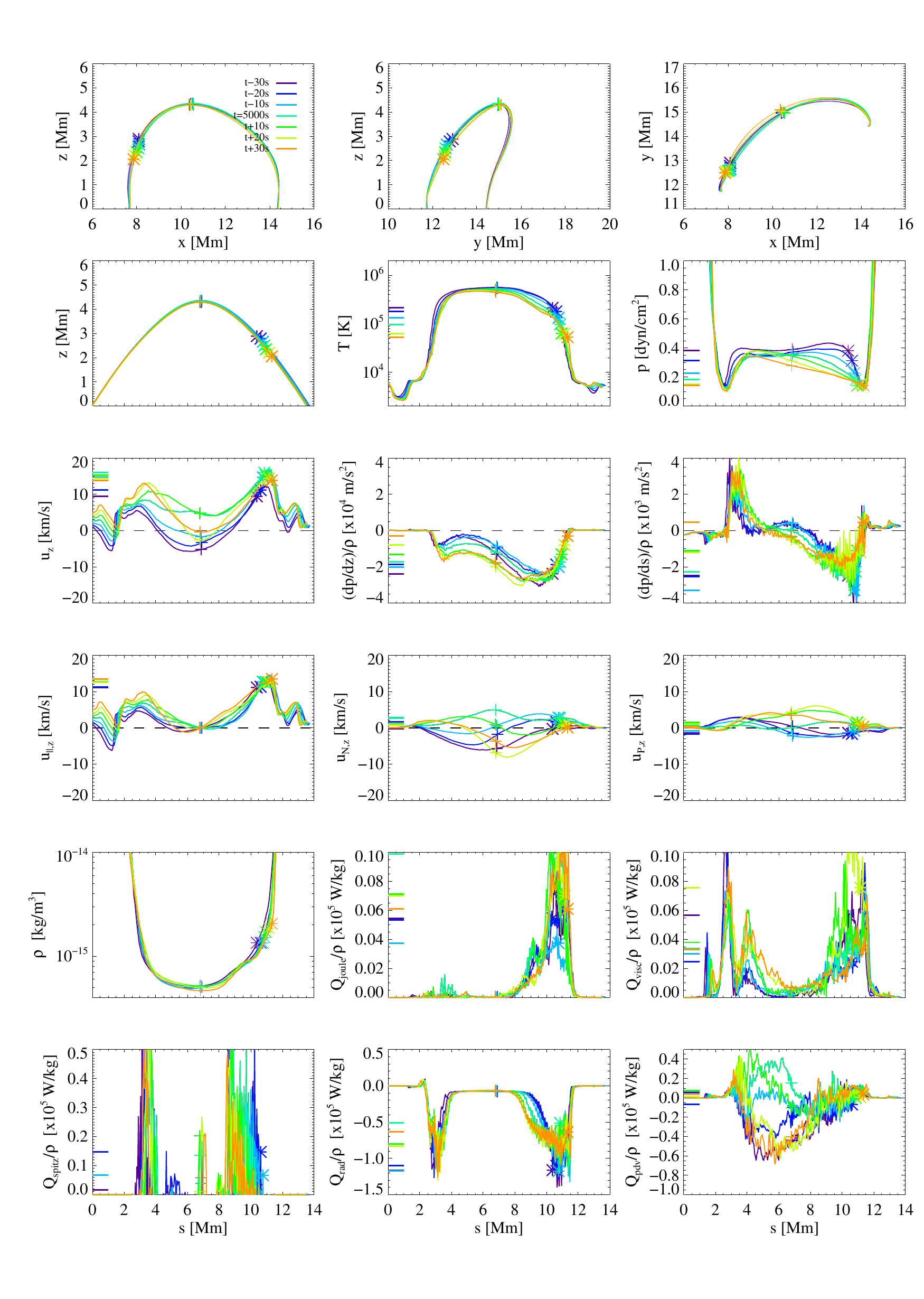}
\end{center}
\caption{Evolution of field line parameters extracted for a low-lying field line including a downward moving transition region cork. The top row shows projections of the field line on the $x$-$z$, $y$-$z,$ and $x$-$y$ planes. The field line parameters are shown as a function of the coordinate $s$ along the field line, where $s$=0~Mm corresponds to the position of the right footpoint of the loop in the panels of the top row. The evolution of the field line parameters is shown for $t=5000\pm$30~s, and the field line evolves from dark blue colors to red colors. The cork positions are indicated by asterisks and the loop apices by crosses.\label{fig11}}
\end{figure*}

In the following, we focus on a bundle of magnetic field lines connected to corks with $T$=100\,000~K, which are showing relatively strong downflows of 15-18 km/s. In Fig. \ref{fig1} (left panel) these field lines are indicated in yellow. The positions of the respective corks and field line apices are indicated by blue and red crosses. The structure we are considering is a rather cool, low-lying loop with an apex height of 4.3~Mm and an apex temperature of $\sim$500\,000~K.
Figure \ref{fig11} shows an example of such a magnetic field line. The top row shows projections of the magnetic field line on the $xz$-, $yz$-, and $xy$-planes for seven consecutive timesteps, {\emph{i.e.,}} $t=5000\pm 30$~s. The evolution of various field line parameters, including temperature, pressure, pressure gradients, velocity components, and various heating and cooling terms, is shown below. 
The magnetic field line does not change much in shape or position over time, and the cork itself is found to be sliding down along the leg of the field line. While moving downward\textbf{}, the cork is cooling from approximately $T$=$200\,000$~K to $T$=$50\,000$~K. 

The central part of the loop has been strongly heated. Since the loop is fairly hot, this heat rapidly spreads throughout the parts of the loop that are hot enough to support significant thermal conduction, leading to a high pressure plug in the upper part of the loop and low pressure dips in regions that have not been heated yet. The details of the two different legs depend on the previous loop history, {\emph{i.e.}}, the initial velocity and density structure, which easily leads to asymmetrical behavior in the two loop legs.

For a more detailed analysis of the (direction of the) flows acting upon the corks, a decomposition of the velocity vector was performed into velocity components parallel ($u_{\parallel}$) and perpendicular to the magnetic field vector ({\it{i.e.,}} $u_{N}$ in the plane of the field line and $u_{P}$ perpendicular to the plane of the field line) at each position along the magnetic field line. These components are given by {$u_{\parallel}={\bf{u}} \cdot {\bf{b}}$, $u_{N}={\bf{u}} \cdot {\bf{N}}$, and $u_{P}={\bf{u}} \cdot {\bf{P}}$, where ${\bf{T}}=\frac{d{\bf{r}}}{ds}={\bf{b}}$, ${\bf{N}}=\frac{d{\bf{T}}}{ds}/|\frac{d{\bf{T}}}{ds}|$, and ${\bf{P}}={\bf{T}}\times {\bf{N}}$ are the unit vectors for a curve ${\bf{r}}(s)$ in 3D space as a function of arc length $s$. In this coordinate system, the vertical velocity $u_z$ can be expressed as the sum of $u_{\parallel ,z} + u_{P,z} + u_{N,z}=u_{\parallel} {\bf{e_{\parallel}}} {\bf{e_z}} + u_P {\bf{e_P}} {\bf{e_z}} + u_{N} {\bf{e_N}} {\bf{e_z}}$.} From a comparison of the parallel and perpendicular velocity components, we can obtain information on the kind of waves and disturbances that are present in the solar atmosphere.

At the position of the cork, the parallel velocity component is found to be stronger than the perpendicular velocity components. For $t=5000$~s, we find $u_{\parallel ,z}$=12.5~km/s, $u_{N,z}$=2.7~km/s, $u_{P,z}$=0.74~km/s, which add up to a total vertical velocity of $u_z$=16.0~km/s. In this case, the position of maximum vertical velocity in the loop is very close to (slightly below) the cork position.

\begin{figure}[!h]
\includegraphics[width=\columnwidth]{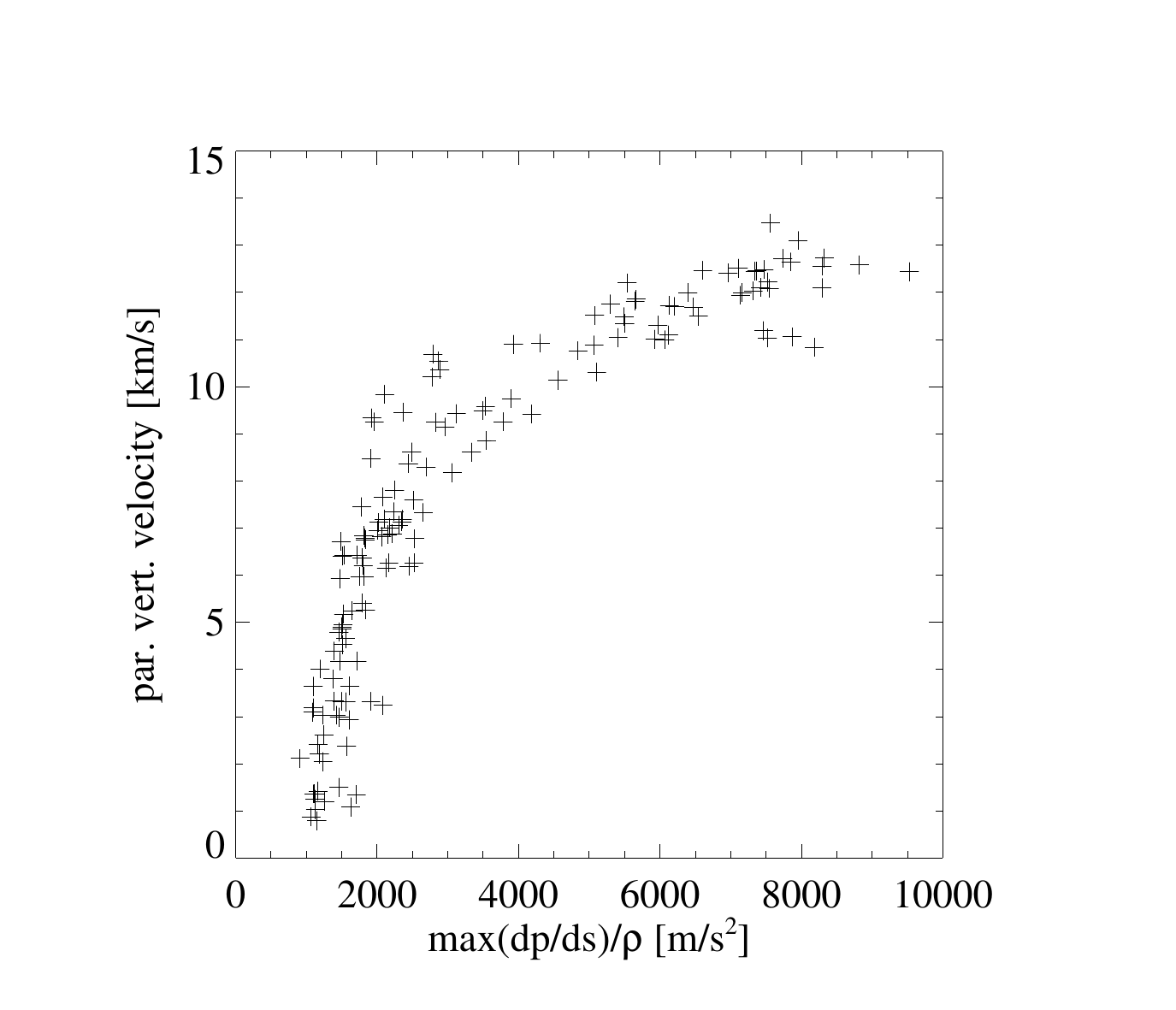}
\caption{Correlation between maximum pressure gradient along the loop and $u_{\parallel,z}$ for a set of downward moving corks on low-lying field lines. The set of magnetic field lines is shown together with the vertical magnetic field strength at height $z$=0~Mm in Fig. \ref{fig1} (left panel). \label{fig12}}
\end{figure}

A closer look at the pressure (gradient) along the field line reveals a higher pressure above the cork than below. The evolution of the plasma pressure shows that the cork is pushed down by a downward propagating pressure front from a high to a low pressure region. This process indicates a correlation between the pressure gradient and the velocity along the loop, where high downflows (redshifts) occur in regions with a large downward directed pressure gradient. Such a correlation is observed for the entire magnetic field line bundle shown in Fig. 1 (see Fig. \ref{fig12}). 
Because of their high density, such low-lying loops are associated with strong radiative losses ($\sim$$\rho^2$). This contributes to the cooling of the corks, as they are pushed down along the field line by a pressure wave from above. The cooling from higher to lower temperatures is typical for corks at transition region temperatures, as described in section \ref{where}. It is not only observed for all corks of the field line bundle shown in Fig. \ref{fig1}, but for many other examples that we investigated. This process is found frequently both in this and other simulations \citep[{\emph{e.g.}},][]{hansteen+al:2010, guerreiro+al:2013}.

\section{Discussion and conclusions}\label{section_discussion}

We have presented the first quantitative study on passive tracer particles, so-called corks, which have been implemented in the Bifrost stellar atmosphere code. In particular, this novel approach sheds new light into mass motions and the formation of redshifts and blueshifts in the solar transition region. Our results show that on timescales of approximately one minute, the majority of corks in the low and middle transition region are cooling and moving downward, while corks in the upper transition region are barely changing temperature and are mostly moving upward. 
The dominance of redshifts in the model can be explained as a result of a higher percentage of downflowing mass compared to upflowing mass, even though the average upflow velocity can be larger than the average downflow velocity in the lower and middle transition region. Blueshifts in the upper transition region can be explained by a combination of both, strong upflows and a higher percentage of upflowing mass compared to downflowing mass in this temperature region. 

It has been shown that at any given time more than half of the corks in the transition region cool rapidly, {\it i.e.}, within a minute, from temperatures above $100\,000$~K to temperatures below $20\,000$~K. Our results are in agreement with the findings of \cite{guerreiro+al:2013} who used tracer fluids at $20\,000$~K and $300\,000$~K in the Bifrost simulations to study mass flows. Whereas \cite{guerreiro+al:2013} speculated that there is rapid heating of the transition region material shortly before it undergoes cooling, our cork based approach has shown that most of the material already has resided at temperatures above $100\,000$~K for several minutes before undergoing cooling.

In this simulation, the cooling process is found to last longer than the heating process and downflows are found to last longer than upflows. The most commonly found situation in the transition region around $100\,000$~K is that corks are cooling and traveling downward. For this set of corks, a strong correlation along magnetic field lines between the pressure gradient and the velocity is found, indicating that a pressure-wave related mechanism is responsible for the observed downflows, {\it i.e.}, material is being pushed downward at high velocity ($\sim$0.3\,$c_s$, where $c_s$ is the speed of sound).

In order to prevent the solar corona from draining, the prevailing downflows in the low and middle transition region must be balanced by upflowing material. This kind of material has been rather difficult to observe, and even less understood is how this material is heated to upper transition region and coronal temperatures on its way upward.
As described in section \ref{section_flows}, our findings indicate that the heating of material from lower to upper transition region and coronal temperatures is a slow process, which takes several minutes, on average, while the material moves through all temperature channels. 
Approximately 4\% of the mass in the temperature interval $T=20\,000-50\,000$~K reaches temperatures above $500\,000$~K within five minutes; 0.5\% reaches coronal temperatures above $800\,000$K within the same time. 
Over a large range of heights, the heating of corks from low to upper transition region temperatures is comparable to the average heating rate in the simulation at the respective height. After having reached upper transition region temperatures, corks tend to remain in this temperature region for several minutes, thus constituting a continuous mass reservoir for the corona.

Previous studies by \cite{hansteen+al:2010} and \cite{guerreiro+al:2013} suggested a common formation mechanism for both observed redshifts and blueshifts based on 
rapid heating scenarios leading to a high pressure plug of material at upper transition region temperatures that relaxes toward equilibrium by expelling material. Pressure wave driven downflows along the field lines as observed in this work complement the picture:\ the larger the downward directed pressure gradient, the higher the observed downflow along the loop (see Fig. \ref{fig12}). The scenario described above (section \ref{field_lines_section}) is a process found frequently both in this simulation and in others. It is a natural consequence of strong heating in a certain region leading to overpressure, forcing material down one or both loop legs.

The cork trajectories were used to follow magnetic field lines over time in a highly accurate way. 
The full exploitation of this approach will be pursued in a follow-up paper, which will focus on the various kinds of loops present in the simulation. In this study, the method was applied to all corks with temperatures of $100\,000$~K at $t=5000$~s. The field lines, which were traced for the transition region corks, have an average apex height of some 5~Mm, indicating that almost all transition region material is found in low-lying and relatively cool loops. 
Treating upflowing and downflowing corks separately did not reveal any differences with respect to the loop height. Using a decomposition of the velocity vector into parallel and perpendicular components along the magnetic field line, we were able to show that for downflowing corks the velocity component parallel to the field is larger than the perpendicular velocity component. 

The validity of this approach is limited by the magnetic Reynolds number $R_m$, {\emph{i.e.}}, the ratio of the advective and diffusive term of the induction equation, which determines how well the assumption of frozen-in magnetic flux holds. Bifrost employs a space- and time-variable nonisotropic magnetic diffusion coefficient $\eta$, and we cannot therefore give a single value for $R_m$. \cite{leenaarts+al:2015} showed that the frozen-in condition holds for a large portion of the simulation domain, however, there are instances in which the field lines evolve independently of the plasma, indicating areas of strong diffusion or reconnection. In this work, we have ignored such field lines.

\section{Outlook}\label{section_outlook}
In this study, we have focused on corks that travel into and out of the solar transition region. 
The corks were implemented into a so-called enhanced network simulation, but many different magnetic field configurations, including flux emergence models, are possible. Their analysis is currently underway. 

We have given a description and first insights into a new method for tracing magnetic flux, which is based on the cork trajectories. The full exploitation of this approach and the analysis of different sets of field lines will be subject of another paper. Future work will also include a more detailed discussion on the various kinds of waves based on the analysis of corks in the Bifrost models.

\begin{acknowledgements}
The authors thank the anonymous referee for his or her thorough review and comments, which contributed to improving the manuscript. The research leading to these results has received funding
from the European Research Council under the European Union's Seventh
Framework Programme (FP7/2007-2013) / ERC Grant agreement No. 291058.
This research was supported by the Research Council of Norway through the
grant "Solar Atmospheric Modelling". 

This research was supported in part with computational resources at NTNU provided by NOTUR, http://www.sigma2.no.
Some computations were performed on resources provided by the Swedish National
 Infrastructure for Computing (SNIC) at the High Performance Computing Center
 North at Ume\aa\ University and the PDC Centre for High Performance Computing
 (PDC-HPC) at the Royal Institute of Technology in Stockholm.

CHIANTI is a collaborative project involving George Mason University, the University of Michigan (USA) and the University of Cambridge (UK). 

\end{acknowledgements}

\bibliography{literature}

\begin{thebibliography}{33}
\expandafter\ifx\csname natexlab\endcsname\relax\def\natexlab#1{#1}\fi

\bibitem[{{Athay}(1984)}]{athay:1984}
{Athay}, R.~G. 1984, \apj, 287, 412

\bibitem[{{Athay} \& {Dere}(1989)}]{athay+dere:1989}
{Athay}, R.~G. \& {Dere}, K.~P. 1989, \apj, 346, 514

\bibitem[{{Athay} \& {Holzer}(1982)}]{athay+holzer:1982}
{Athay}, R.~G. \& {Holzer}, T.~E. 1982, \apj, 255, 743

\bibitem[{{Bingert} \& {Peter}(2011)}]{bingert+peter:2011}
{Bingert}, S. \& {Peter}, H. 2011, \aap, 530, A112

\bibitem[{{Brekke}(1993)}]{brekke:1993}
{Brekke}, P. 1993, \apj, 408, 735

\bibitem[{{Brekke} {et~al.}(1997){Brekke}, {Hassler}, \&
  {Wilhelm}}]{brekke+al:1997}
{Brekke}, P., {Hassler}, D.~M., \& {Wilhelm}, K. 1997, \solphys, 175, 349

\bibitem[{{Carlsson} {et~al.}(2016){Carlsson}, {Hansteen}, {Gudiksen},
  {Leenaarts}, \& {De Pontieu}}]{carlsson+al:2016}
{Carlsson}, M., {Hansteen}, V.~H., {Gudiksen}, B.~V., {Leenaarts}, J., \& {De
  Pontieu}, B. 2016, \aap, 585, A4

\bibitem[{{Carlsson} \& {Leenaarts}(2012)}]{carlsson+leenaarts:2012}
{Carlsson}, M. \& {Leenaarts}, J. 2012, \aap, 539, A39

\bibitem[{{Chae} {et~al.}(1998){Chae}, {Yun}, \& {Poland}}]{chae+al:1998}
{Chae}, J., {Yun}, H.~S., \& {Poland}, A.~I. 1998, \apjs, 114, 151

\bibitem[{{Dadashi} {et~al.}(2011){Dadashi}, {Teriaca}, \&
  {Solanki}}]{dadashi+al:2011}
{Dadashi}, N., {Teriaca}, L., \& {Solanki}, S.~K. 2011, \aap, 534, A90

\bibitem[{{Dere} {et~al.}(1984){Dere}, {Bartoe}, \& {Brueckner}}]{dere+al:1984}
{Dere}, K.~P., {Bartoe}, J.-D.~F., \& {Brueckner}, G.~E. 1984, \apj, 281, 870

\bibitem[{{Dere} {et~al.}(1989){Dere}, {Bartoe}, {Brueckner}, \&
  {Recely}}]{dere+al:1989}
{Dere}, K.~P., {Bartoe}, J.-D.~F., {Brueckner}, G.~E., \& {Recely}, F. 1989,
  \apjl, 345, L95

\bibitem[{{Doschek} {et~al.}(1976){Doschek}, {Bohlin}, \&
  {Feldman}}]{doschek+al:1976}
{Doschek}, G.~A., {Bohlin}, J.~D., \& {Feldman}, U. 1976, \apjl, 205, L177

\bibitem[{{Feldman} {et~al.}(1982){Feldman}, {Doschek}, \&
  {Cohen}}]{feldman+al:1982}
{Feldman}, U., {Doschek}, G.~A., \& {Cohen}, L. 1982, \apj, 255, 325

\bibitem[{{Gebbie} {et~al.}(1981){Gebbie}, {Hill}, {November}, {Gurman},
  {Shine}, {Woodgate}, {Athay}, {Tandberg-Hanssen}, {Toomre}, \&
  {Simon}}]{gebbie+al:1981}
{Gebbie}, K.~B., {Hill}, F., {November}, L.~J., {et~al.} 1981, \apjl, 251, L115

\bibitem[{{Gudiksen} {et~al.}(2011){Gudiksen}, {Carlsson}, {Hansteen}, {Hayek},
  {Leenaarts}, \& {Mart{\'{\i}}nez-Sykora}}]{gudiksen+al:2011}
{Gudiksen}, B.~V., {Carlsson}, M., {Hansteen}, V.~H., {et~al.} 2011, \aap, 531,
  A154

\bibitem[{{Gudiksen} \& {Nordlund}(2005)}]{gudiksen+nordlund:2005}
{Gudiksen}, B.~V. \& {Nordlund}, {\AA}. 2005, \apj, 618, 1020

\bibitem[{{Guerreiro} {et~al.}(2013){Guerreiro}, {Hansteen}, \& {De
  Pontieu}}]{guerreiro+al:2013}
{Guerreiro}, N., {Hansteen}, V., \& {De Pontieu}, B. 2013, \apj, 769, 47

\bibitem[{{Hansteen}(1993)}]{hansteen:1993}
{Hansteen}, V. 1993, \apj, 402, 741

\bibitem[{{Hansteen} {et~al.}(2015){Hansteen}, {Guerreiro}, {De Pontieu}, \&
  {Carlsson}}]{hansteen+al:2015}
{Hansteen}, V., {Guerreiro}, N., {De Pontieu}, B., \& {Carlsson}, M. 2015,
  \apj, 811, 106

\bibitem[{{Hansteen} {et~al.}(2007){Hansteen}, {Carlsson}, \&
  {Gudiksen}}]{hansteen+carlsson+gudiksen:2007}
{Hansteen}, V.~H., {Carlsson}, M., \& {Gudiksen}, B. 2007, in Astronomical
  Society of the Pacific Conference Series, Vol. 368, The Physics of
  Chromospheric Plasmas, ed. P.~{Heinzel}, I.~{Dorotovi{\v c}}, \& R.~J.
  {Rutten}, Heinzel

\bibitem[{{Hansteen} {et~al.}(2010){Hansteen}, {Hara}, {De Pontieu}, \&
  {Carlsson}}]{hansteen+al:2010}
{Hansteen}, V.~H., {Hara}, H., {De Pontieu}, B., \& {Carlsson}, M. 2010, \apj,
  718, 1070

\bibitem[{Hyman(1979)}]{hyman:1979}
Hyman, J.~M. 1979, in Proceedings of the third IMACS international symposium on
  computer methods for partial differential equations, ed. R.~Vichnevetsky \&
  R.~Stepleman, International Association for Mathematics and Computers in
  Simulation, Dept. of Computer Science, Rutgers University, New Brunswick,
  N.J. 08903 USA, IMACS: 313,3

\bibitem[{{Klimchuk}(1987)}]{klimchuk:1987}
{Klimchuk}, J.~A. 1987, \apj, 323, 368

\bibitem[{{Leenaarts} {et~al.}(2015){Leenaarts}, {Carlsson}, \& {Rouppe van der
  Voort}}]{leenaarts+al:2015}
{Leenaarts}, J., {Carlsson}, M., \& {Rouppe van der Voort}, L. 2015, \apj, 802,
  136

\bibitem[{{Mariska} {et~al.}(1978){Mariska}, {Feldman}, \&
  {Doschek}}]{mariska+al:1978}
{Mariska}, J.~T., {Feldman}, U., \& {Doschek}, G.~A. 1978, \apj, 226, 698

\bibitem[{{Peter}(1999)}]{peter:1999}
{Peter}, H. 1999, \apj, 516, 490

\bibitem[{{Peter} {et~al.}(2006){Peter}, {Gudiksen}, \&
  {Nordlund}}]{peter+al:2006}
{Peter}, H., {Gudiksen}, B.~V., \& {Nordlund}, {\AA}. 2006, \apj, 638, 1086

\bibitem[{{Peter} \& {Judge}(1999)}]{peter+judge:1999}
{Peter}, H. \& {Judge}, P.~G. 1999, \apj, 522, 1148

\bibitem[{{Pneuman} \& {Kopp}(1977)}]{pneuman+kopp:1977}
{Pneuman}, G.~W. \& {Kopp}, R.~A. 1977, \aap, 55, 305

\bibitem[{{Rottman} {et~al.}(1990){Rottman}, {Hassler}, {Jones}, \&
  {Orrall}}]{rottman+al:1990}
{Rottman}, G.~J., {Hassler}, D.~D., {Jones}, M.~D., \& {Orrall}, F.~Q. 1990,
  \apj, 358, 693

\bibitem[{{Spitzer}(1956)}]{spitzer:1956}
{Spitzer}, L. 1956, {Physics of Fully Ionized Gases} (Interscience Publishers,
  New York)

\bibitem[{Zacharias {et~al.}(2011)Zacharias, Peter, \&
  Bingert}]{zacharias+al:2011a}
Zacharias, P., Peter, H., \& Bingert, S. 2011, Astronomy {\&} Astrophysics,
  531, 97

\end{thebibliography}
\bibliographystyle{aa}

\end{document}